\documentclass[jcp,aip,10pt,showpacs,reprint,floatfix]{revtex4-1}

\usepackage{amsmath,amssymb,amsfonts,graphicx}  

\newcommand{\be}{\begin{equation}}
\newcommand{\ee}{\end{equation}}
\newcommand{\bp}{\begin{picture}}
\newcommand{\ep}{\end{picture}}
\newcommand{\ba}[1]{\begin{array}{#1}}
\newcommand{\ea}{\end{array}}
\newcommand{\bea}{\begin{eqnarray}}
\newcommand{\eea}{\end{eqnarray}}

\newcommand{\LV}{liquid-vapor }
\newcommand{\eq}[1]{Eq.~(\ref{#1})}
\newcommand{\fig}[1]{Fig.~\ref{#1}}

\newcommand{\sect}[1]{Section~\ref{#1}}
\newcommand{\avg}[1]{\langle #1 \rangle}
\newcommand{\olcite}[1]{Ref.~\onlinecite{#1}}
\newcommand{\ahum}[1]{``#1''}

\newcommand{\NC}{{N_{\rm c}}}
\newcommand{\NP}{N_{\rm p}}
\newcommand{\etac}{\eta_{\rm c}}
\newcommand{\etapr}{\eta_{\rm p}^{\rm r}}
\newcommand{\rhoc}{\rho_{\rm c}}
\newcommand{\rhop}{\rho_{\rm p}}
\newcommand{\etaprtr}{\eta_{\rm p,tr}^{\rm r}}
\newcommand{\etaprcrvz}{\eta_{\rm p,cr}^{\rm r,vz}}
\newcommand{\etaprcrlz}{\eta_{\rm p,cr}^{\rm r,lz}}
\newcommand{\etaprcrbulk}{\eta_{\rm p,cr}^{\rm r,bulk}}
\newcommand{\muc}{\mu_{\rm c}}
\newcommand{\mucvz}{\mu_{\rm c}^{\rm vz}}
\newcommand{\mucvzcr}{\mu_{\rm c,cr}^{\rm vz}}
\newcommand{\muclz}{\mu_{\rm c}^{\rm lz}}
\newcommand{\muclzcr}{\mu_{\rm c,cr}^{\rm lz}}
\newcommand{\mup}{\mu_{\rm p}}

\begin{document}

\title{Fluid phase separation inside a static periodic field: an effectively \\ 
two-dimensional critical phenomenon}

\author{Richard L. C. Vink}
\affiliation{Institute of Theoretical Physics, Georg-August-Universit\"at 
G\"ottingen, Friedrich-Hund-Platz~1, 37077 G\"ottingen, Germany}

\author{Tim Neuhaus} 
\author{Hartmut L\"owen} 
\affiliation{Institut f\"{u}r Theoretische Physik II: Weiche Materie, 
Heinrich-Heine-Universit\"{a}t D\"usseldorf, Universit\"{a}tsstra{\ss}e 1, 40225 
D\"{u}sseldorf, Germany}

\date{\today}

\begin{abstract} When a fluid with a bulk \LV critical point is placed inside a 
static external field with spatial periodic oscillations in one direction, the 
bulk critical point splits into two new critical points and a triple point. This 
phenomenon is called laser-induced condensation [Mol.~Phys.~{\bf 101}, 1651 
(2003)], and it occurs when the wavelength of the field is sufficiently large. 
The critical points mark the end of two coexistence regions, namely between (1) 
a vapor and stacked-fluid phase, and (2) a stacked-fluid and liquid phase. The 
stacked-fluid or \ahum{zebra} phase is characterized by large density 
oscillations along the field direction. We study the above phenomenon for a 
mixture of colloids and polymers using density functional theory and computer 
simulation. The theory predicts that the vapor-zebra and liquid-zebra surface 
tensions are extremely small. Most strikingly, however, is the theoretical 
finding that at their respective critical points, both tensions vanish, but not 
according to any critical power law. The solution to this apparent paradox is 
provided by the simulations. These show that the field divides the system into 
effectively two-dimensional slabs, stacked on top of each other along the field 
direction. Inside each slab, the system behaves as if it were two-dimensional, 
while in the field direction the system resembles a one-dimensional Ising chain. 
\end{abstract}

\pacs{61.20.Gy,68.05.-n,82.70.Dd}


\maketitle

\section{Introduction}

Binary mixtures of sterically-stabilized colloids and non-adsorbing globular 
polymers are valuable model systems. In particular, the addition of polymers 
causes an effective depletion attraction between the colloids. This can induce 
\LV type transitions in these systems, in much the same way as in an atomic 
fluid. Indeed, \LV demixing in colloid-polymer mixtures has been routinely 
studied in the past, using theory, computer simulation, and experiment 
\cite{Brader2003, Lekkerkerker1992, Poon2002, Tuinier2008, Bolhuis2002, 
Dzubiella2001}. More phenomena for which colloid-polymer mixtures are ideal 
model systems include equilibrium clustering \cite{Stradner2004}, 
\ahum{attractive} glasses \cite{Pham2002, Zaccarelli2004}, gelation 
\cite{Laurati2009}, numerous interfacial phenomena \cite{Vliegenthart1997, 
Hoog1999, Chen2000, Chen2001} including capillary waves \cite{Aarts2004, 
Vink2005}, and wetting \cite{Aarts2003, Wijting2003, Wijting2003a, Indekeu2010}.

Presumably the simplest model of a colloid-polymer mixture is the one proposed 
by Asakura and Oosawa (AO) \cite{Asakura1954, Asakura1958, Vrij1976}. Despite 
its simplicity, this model captures the essential physics \cite{Poon2002, 
Bolhuis2002}, yet remains simple enough to allow for both theoretical 
investigations (based, for instance, on liquid state and density functional 
theory \cite{Schmidt2000, Schmidt2002}), and computer simulations 
\cite{Vink2004, Dijkstra2002}. In agreement with experiments, the AO model 
features a bulk \LV critical point \cite{Lekkerkerker1992}, a freezing 
transition \cite{Zykova-Timan2010}, and also confinement effects at a single 
wall \cite{Wessels2004, Wessels2004a}, or between two parallel walls 
\cite{Binder2008, DeVirgiliis2008, DeVirgiliis2007, Vink2006a, Fortini2006, 
Schmidt2003} can be studied using this model. The AO model has also been used to 
study wetting \cite{Brader2000, Evans2001, Brader2002}, as well as phase 
separation in porous media \cite{Wessels2003, Wessels2005, Vink2006, 
Pellicane2008, Vink2008, Vink2009}. Also of interest is the phase behavior of 
the AO model inside an external field, such as gravity \cite{Schmidt2004, 
Jamie2010}, or a spatially-varying periodic field \cite{Gotze2003}.

In \olcite{Gotze2003}, a colloid-polymer mixture described by the AO model was 
subjected to a standing-wave external field, wavelength $\lambda$, propagating 
in one direction (the $z$-direction in what follows). The field is thus 
effectively one-dimensional. This can be realized in experiments by placing a 
colloidal suspension inside a standing laser field \cite{Freire1994}. While the 
influence of such a field on the {\it freezing} transition has been extensively 
studied, and is known to induce laser-induced freezing \cite{Chaudhuri2005, 
Franzrahe2009}, relatively little is known regarding its effect on the \LV 
transition. Regarding the latter, \olcite{Gotze2003} proposes the following 
scenario: for sufficiently large $\lambda$, the bulk \LV critical point splits 
into two critical points and a triple point with an intermediate new phase which 
is partially condensed in slabs perpendicular to the $z$-direction. We call this 
phase the \ahum{zebra} phase in what follows. The results of \olcite{Gotze2003} 
were obtained by using fundamental measure density functional theory for a 
colloid-polymer mixture described by the AO model.

In this work, we use the same model and technique to calculate the surface 
tensions between all three coexisting phases. We find that the vapor-zebra and 
liquid-zebra surface tensions are extremely small. Moreover, to our surprise, 
upon approach of the critical points, the latter tensions do not yield the 
expected critical power law behavior. To clarify the nature of the critical 
points, we use Monte Carlo simulations and finite-size scaling. The simulations 
confirm all the trends predicted by the theory, and also explain why the 
vapor-zebra and liquid-zebra surface tensions do not become critical. The main 
finding is that the standing-wave external field divides the system into 
effectively two-dimensional slabs, stacked on top of each other along the 
$z$-direction. Inside each slab, the system behaves as if it were 
two-dimensional, while in the $z$-direction the system resembles an effectively 
one-dimensional system. Hence, by turning on the external field, the bulk system 
is \ahum{split-up} into $2+1$ separate dimensions. With this picture in mind, 
most of our observations can be intuitively understood.

\section{Model and unit conventions}
\label{sec:model}

To describe the interactions between colloids (c) and polymers (p) we use the 
Asakura-Oosawa (AO) model \cite{Asakura1954, Asakura1958, Vrij1976}. The 
colloids and the polymer coils are both assumed to be spherical objects with 
respective diameters $\sigma$ and $\sigma_{\rm p}$. In what follows, the colloid 
diameter $\sigma \equiv 1$ will be the unit of length, and the 
colloid-to-polymer size ratio is denoted $q=\sigma_{\rm p}/\sigma$. The 
interaction between colloid-colloid and colloid-polymer pairs is hard-core, 
while the polymer-polymer interaction is ideal, leading to the following pair 
potentials
\begin{eqnarray}
u_{\rm cc}(r) &=& \begin{cases}
 \infty & r<\sigma \\
 0 & \mbox{otherwise,}
\end{cases} \\
u_{\rm cp}(r) &=& \begin{cases}
 \infty & r< (\sigma+\sigma_{\rm p})/2 \\
 0 & \mbox{otherwise,} 
\end{cases} \\
u_{\rm pp}(r) &=& 0,
\end{eqnarray}
with $r$ the center-to-center distance. As the interactions are either hard-core 
or ideal, the temperature~$T$ can be scaled out (it only sets the energy scale 
$k_BT$, where $k_B$ is the Boltzmann constant). We mostly use a grand canonical 
ensemble, i.e.~the system volume $V$, the colloid chemical potential $\muc$, and 
the polymer chemical potential $\mu_{\rm p}$ are fixed, but the number of 
colloids $\NC$ and polymers $\NP$ inside $V$ fluctuates. The particle densities 
are defined as $\rho_i = N_i/V$, with $i \in (\rm c,p)$. We also introduce the 
colloid packing fraction $\etac = \rho_{\rm c} v_{\rm c}$, with $v_{\rm c} = \pi 
\sigma^3/6$ the volume of a single colloid. Following convention 
\cite{Lekkerkerker1992}, we do not use the polymer chemical potential itself, 
but rather the polymer reservoir packing fraction $\etapr$. It is defined as the 
packing fraction $\etapr \equiv \rho_{\rm p} v_{\rm p}$, $v_{\rm p} = \pi 
\sigma_{\rm p}^3/6$, of a {\it pure} polymer system at given $\mu_{\rm p}$ (for 
the AO model, such a system is simply an ideal gas, and hence $\etapr \propto 
e^{\mu_{\rm p} / k_BT}$). In order to compare $\muc$ between theory and 
simulation, the thermal wavelength $\Lambda=\sigma/2$ in what follows.

In the AO model, there is an effective attraction between the colloids. This can 
be shown formally by \ahum{integrating out} the polymers \cite{Asakura1954, 
Dijkstra1999}, which leads to an effective colloid-colloid pair potential with 
an attractive well; the well-depth is proportional to $\etapr$. Hence, the {\it 
bulk} AO model undergoes a \LV type transition, with $\etapr$ playing the role 
of inverse temperature (by bulk we explicitly mean a three-dimensional system in 
the absence of any surfaces or fields).

The extension of this work is to consider the AO model inside an external field 
propagating along the $z$-direction
\begin{equation}\label{eq:laser}
 V_{\rm ext}(z) = V_0 \cos \left( 2\pi z / \lambda \right),
\end{equation}
with $V_0$ the field amplitude, and $\lambda$ the wavelength. We emphasize that 
the field is static: it oscillates in space, not in time. Physically, 
\eq{eq:laser} resembles a one-dimensional standing optical wave, which could 
easily be realized experimentally using a laser beam. In this work, we assume 
that \eq{eq:laser} acts on the colloidal particles only, but that the polymers 
do not \ahum{feel} the field. The Hamiltonian of the system is thus defined by 
the AO pair potentials, Eqs.(1)-(3), plus an external field contribution 
$\sum_{i=1}^\NC V_{\rm ext}(z_i)$, where the sum is over all colloids, and with 
$z_i$ the $z$-coordinate of the $i$-th colloidal particle.

\section{Density functional theory}

In density functional theory (DFT), the equilibrium density profiles are the 
ones that minimize the grand canonical free energy functional
\be\label{eq:func}
 \Omega[T, \muc, \etapr, \rho_{\rm c}(z), \rho_{\rm p}(z)],
\ee
where $\rho_{\rm c}(z)$ is the average colloid density at position $z$ along the 
field direction, and $\rho_{\rm p}(z)$ that of the polymers (we thus use an 
effectively one-dimensional set-up in our DFT calculations). In what follows, we 
also use the colloid packing fraction profile $\etac(z)$, which is simply 
$\rho_{\rm c}(z)$ multiplied by $v_{\rm c}$. The system is symmetric 
around~$z=0$, and we include periodic boundary conditions. Based on the proof 
that a free energy functional $\Omega$ indeed exists \cite{Mermin1965}, we use 
the fundamental measure approach of \olcite{Schmidt2000} to approximate 
\eq{eq:func}; see Appendix~\ref{app:dft} for full details. We numerically solve 
the resulting stationarity equation using a Picard iteration scheme 
\cite{Roth2010}.

\subsection{Phase diagram}

\begin{figure}
\centering
\includegraphics[width=0.9\columnwidth]{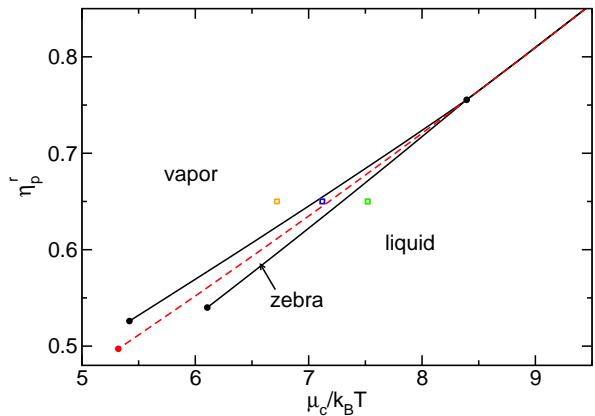}
\caption{\label{fig:phase-diagram_GOETZE} Phase diagram of the AO model with 
$q=0.6$ in $(\muc,\etapr)$~representation as obtained using DFT. The dashed 
curve shows the bulk binodal, the solid curves the binodals inside the external 
field of \eq{eq:laser} using $\lambda=8.192$, and $V_0 = 0.5 \, k_BT$. The lower 
three dots indicate critical points, the upper dot the triple point. The open 
squares denote the state points at which the colloid density profiles of 
\fig{fig:gas-fluid-zebra_GOETZE} were measured.}
\end{figure}

\begin{figure}
\centering
\includegraphics[width=0.9\columnwidth]{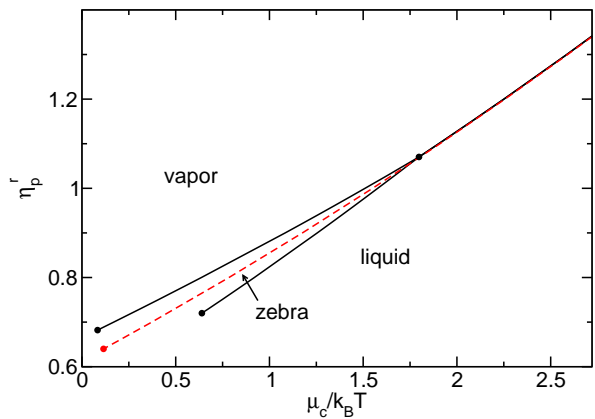}
\caption{\label{fig:phase-diagram_RVPAR} Same as \fig{fig:phase-diagram_GOETZE}, 
but using $q=1.0$, with external field parameters $\lambda=10$, and $V_0=0.4 \, 
k_BT$.}
\end{figure}

First, we revisit \olcite{Gotze2003} and hence choose the size ratio $q=0.6$, 
the wavelength of the external field $\lambda=8.192$, and its amplitude $V_0=0.5 
\, k_BT$. \fig{fig:phase-diagram_GOETZE} shows the phase diagram obtained from 
our DFT calculation. Here, we use the grand canonical representation, i.e.~we 
plot the binodals in the $(\muc, \etapr)$~plane. Clearly visible is the 
characteristic \ahum{inverted letter Y} or \ahum{pitchfork} topology (solid 
lines). For comparison, the dashed line shows the bulk binodal, i.e.~obtained 
without the external potential of \eq{eq:laser}. We first note that the bulk 
critical point occurs at a value of $\etapr$ below that of the vapor-zebra and 
liquid-zebra critical points. This is to be expected as confinement generally 
lowers transition temperatures. The more striking feature of 
\fig{fig:phase-diagram_GOETZE} is that $\etapr$ of the liquid-zebra critical 
point exceeds that of the vapor-zebra one: we find $\etaprcrvz \approx 0.526$ 
and $\etaprcrlz \approx 0.540$. In contrast, in \olcite{Gotze2003}, no 
difference could be detected, which demonstrates the improved accuracy of the 
present work. The fact that the critical \ahum{inverse temperatures} differ is a 
genuine feature, since our simulations reveal the same trend. As $\etapr$ 
increases, the binodals approach each other, and at the triple point, $\etaprtr 
\approx 0.755$, they meet. In order to facilitate the comparison to computer 
simulation later on, we also present the phase diagram for $q=1$, using field 
parameters $\lambda=10$, and $V_0=0.4 \, k_BT$ (\fig{fig:phase-diagram_RVPAR}). 
We obtain the same overall topology, but the region where the zebra phase occurs 
has broadened. In addition, the binodals are shifted to significantly lower 
colloid chemical potential.

\begin{figure}
\centering
\includegraphics[width=0.9\columnwidth]{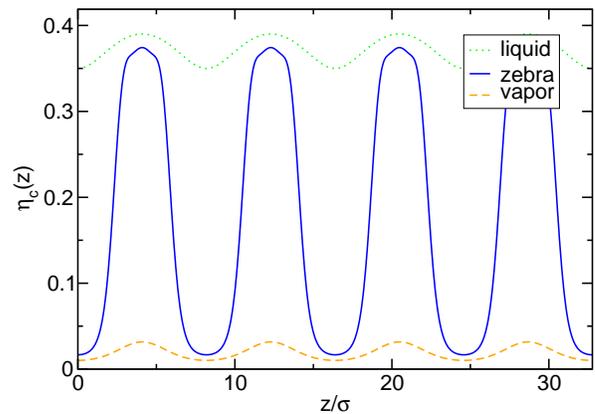}
\caption{\label{fig:gas-fluid-zebra_GOETZE} Equilibrium colloid density profiles 
$\etac(z)$ measured along the direction of the laser field for the three state 
points indicated by open squares in the phase diagram of 
\fig{fig:phase-diagram_GOETZE}, corresponding to the vapor phase 
($\muc/k_BT=6.72$), the zebra phase ($\muc/k_BT=7.12$), and the liquid phase 
($\muc/k_BT=7.52$). The profiles were obtained at fixed $\etapr=0.65$.}
\end{figure}

Next, we consider the structural properties of the phases. The key difference 
with the bulk AO model is that, in addition to a vapor and liquid phase, we now 
also have the zebra phase. The latter phase arises when $\etapr$ is chosen 
between the critical and triple points, and with the colloid chemical potential 
chosen appropriately. To characterize the phases, we have measured colloid 
density profiles $\etac(z)$ along the direction of the laser field at three 
points in the phase diagram of \fig{fig:phase-diagram_GOETZE}, indicated by open 
squares. The latter correspond, from left to right, to a vapor state, a zebra 
state, and a liquid state. The density profiles are shown in 
\fig{fig:gas-fluid-zebra_GOETZE}. The salient feature is that all three phases 
display density modulations in the $z$-direction, but the average density and 
amplitude differ. The average colloid density is low in the vapor phase, high 
in the liquid phase, with only a modest density amplitude in both phases. The 
most striking feature of the zebra phase is the unusually large density 
amplitude, which oscillates between the average density of the vapor and liquid 
phase.

\subsection{Interfaces and interfacial free energies}
\label{sec:interfaces}

\begin{figure}
\centering
\includegraphics[width=0.9\columnwidth]{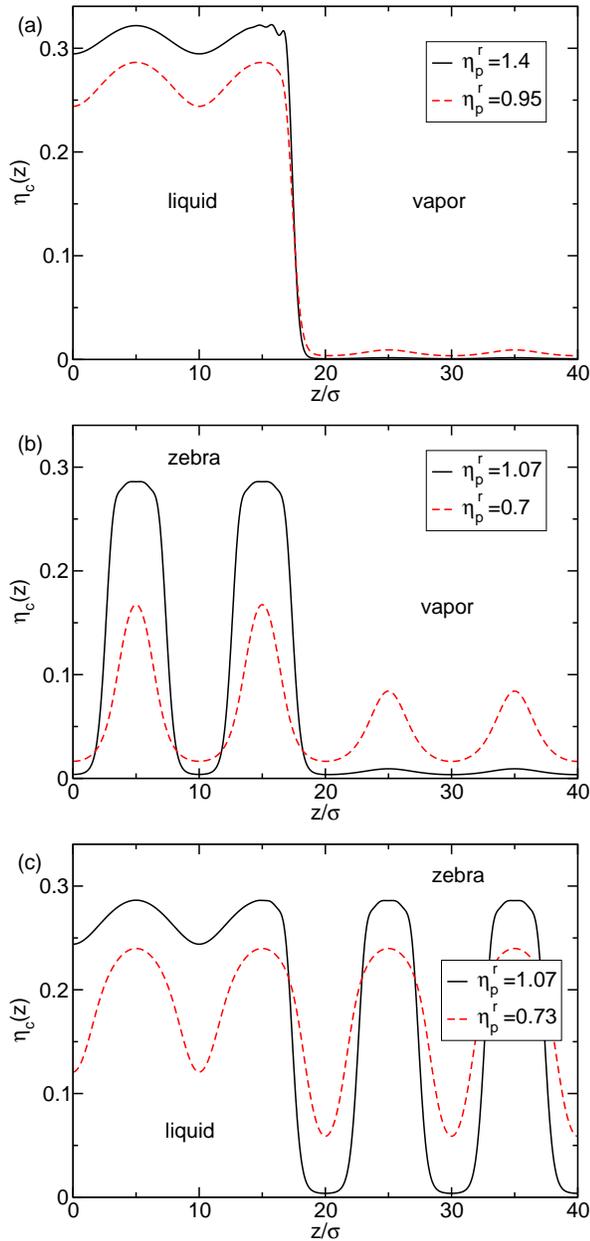}
\caption{\label{fig:dft_cx} Equilibrium colloid density profiles $\etac(z)$ 
showing the various interfaces for the AO model with $q=1$, inside the external 
field of \eq{eq:laser}, using field parameters $\lambda=10$, and $V_0 = 0.4 \, 
k_B T$. Shown is the \LV interface (a), the vapor-zebra interface (b), 
and the liquid-zebra interface (c), each time for two values of $\etapr$ as 
indicated.}
\end{figure}

We now consider the interfaces between the coexisting phases and the 
corresponding surface tensions. Above the triple point, $\etapr>\etaprtr$, 
liquid and vapor coexist, with a corresponding liquid-vapor surface tension 
$\gamma_{\rm lv}$. To calculate $\gamma_{\rm lv}$, we first compute the 
equilibrium colloid density profiles $\etac(z)$ of the pure vapor and liquid 
phase; the latter yield the Gibbs free energies $\Omega_{\rm v,pure}$ and 
$\Omega_{\rm l,pure}$, respectively (at coexistence: $\Omega_{\rm v,pure} = 
\Omega_{\rm l,pure} \equiv \Omega_{\rm pure}$). The density profiles of the pure 
phases schematically resemble those of \fig{fig:gas-fluid-zebra_GOETZE}. Next, 
we consider $\etac(z)$ of a system containing a liquid-vapor interface, from 
which we obtain $\Omega_{\rm lv,int}$. An example is shown in 
\fig{fig:dft_cx}(a), using two values of $\etapr$. For sufficiently large 
$\etapr$, we observe small oscillations at high densities close to the 
interface. Since the surface tension is the excess free energy per unit of area, 
it follows that
\begin{equation}\label{eq:dft_gamma}
 \gamma_{\rm lv} = \frac{\Omega_{\rm pure} - \Omega_{\rm lv,int}}{2A},
\end{equation}
where $A$ is the area of the interface (the factor $1/2$ results from the fact 
that two interfaces are present in our DFT set-up). The calculation of the 
vapor-zebra surface tension $\gamma_{\rm vz}$, and of the liquid-zebra surface 
tension $\gamma_{\rm lz}$, which become defined below the triple point, is 
performed analogously. To this end, one needs to compute $\etac(z)$ for a system 
containing a vapor-zebra and liquid-zebra interface. Some typical density 
profiles are shown in \fig{fig:dft_cx}(b) and~(c). It is striking that the 
interfaces are extremely sharp: even very close to the interface position, the 
density profiles of the coexisting phases are almost identical to those obtained 
without the interface! From this observation, and comparing to 
\eq{eq:dft_gamma}, one can already deduce that $\gamma_{\rm vz}, \gamma_{\rm 
lz}$ must be very small.

\begin{figure}
\centering
\includegraphics[width=0.9\columnwidth]{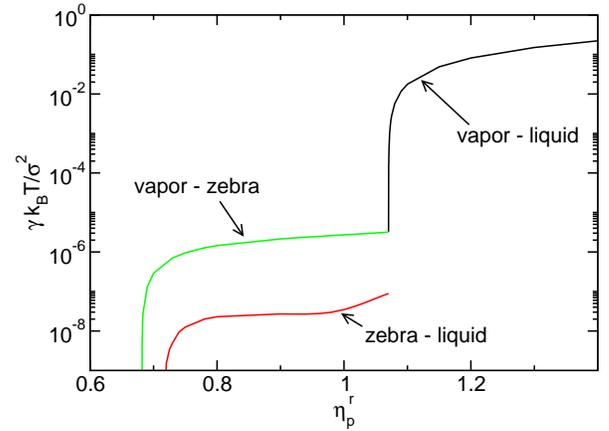}
\caption{\label{fig:tension_RVPAR} Variation of the surface tensions 
$\gamma_{\rm lv}$, $\gamma_{\rm vz}$, and $\gamma_{\rm lz}$ with $\etapr$. Note 
in particular the extremely small values of $\gamma_{\rm vz}$ and $\gamma_{\rm 
lz}$. These data were obtained using our DFT for the AO model with $q=1$, and 
external field parameters $\lambda=10$, $V_0 = 0.4 \, k_BT$.}
\end{figure}

In \fig{fig:tension_RVPAR}, we summarize the results of the DFT surface tension 
calculations, where the various tensions are plotted as function of $\etapr$. 
Starting above the triple point, $\gamma_{\rm lv}$ is finite; by decreasing 
$\etapr$, $\gamma_{\rm lv}$ vanishes at the triple point. At the triple point, 
$\gamma_{\rm vz}$ and $\gamma_{\rm lz}$ are finite; by decreasing $\etapr$ 
further, $\gamma_{\rm vz}$ vanishes at $\etapr=\etaprcrvz$ of the vapor-zebra 
critical point, while $\gamma_{\rm lz}$ vanishes at $\etapr=\etaprcrlz$ of the 
liquid-zebra critical point. Note the extremely small values of $\gamma_{\rm 
vz}$ and $\gamma_{\rm lz}$ over the entire range between the critical and triple 
points. In fact, it always holds that $\gamma_{\rm lv} \geq \gamma_{\rm 
vz}+\gamma_{\rm lz}$, which implies there is no complete wetting of the 
\ahum{zebra} phase for $\etapr>\etaprtr$.

\subsection{Critical behavior}

\begin{figure}
\centering
\includegraphics[width=0.9\columnwidth]{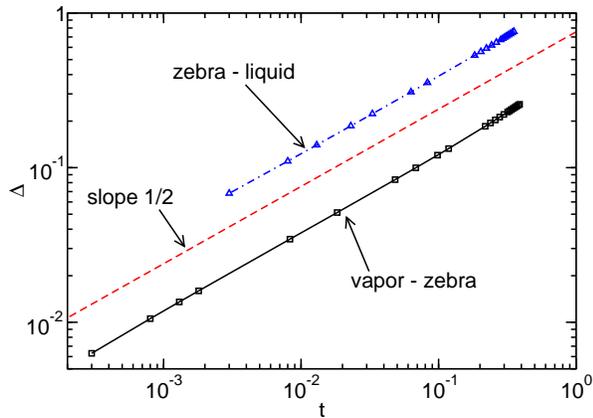}
\caption{\label{fig:beta_RVPAR} Order parameters $\Delta_{\rm vz}$ and 
$\Delta_{\rm lz}$ as function of the distance $t$ from their respective critical 
points, where $\etaprcrvz \approx 0.6817$ and $\etaprcrlz \approx 0.717$ were 
used. Note the double logarithmic scales! The straight line corresponds to a 
power law with critical exponent $\beta=1/2$ of mean-field theory. For clarity, 
the data for $\Delta_{\rm lz}$ have been shifted upward by half a decade.}
\end{figure}

\begin{figure}
\centering
\includegraphics[width=0.9\columnwidth]{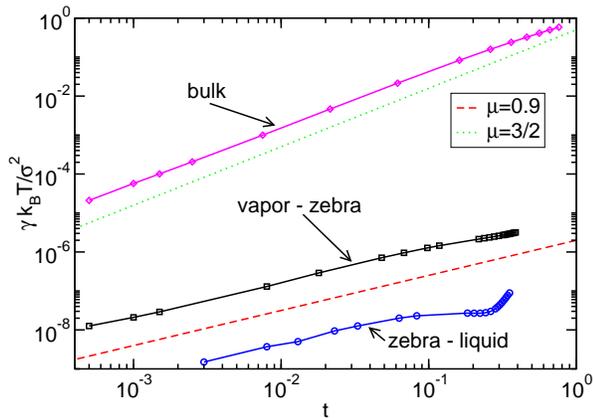}
\caption{\label{fig:dft_st} Surface tensions $\gamma_{\rm vz}$, $\gamma_{\rm 
lz}$, and the {\it bulk} \LV surface tension, as function of the distance~$t$ 
from the respective critical points, where for $\etaprcrvz$ and $\etaprcrlz$ the 
values of \fig{fig:beta_RVPAR} were used, and $\etaprcrbulk \approx 0.6385$. 
Note again the double logarithmic scales! The straight line corresponds to a 
power law with critical exponent $\mu=3/2$ of mean-field theory. The key message 
is that $\gamma_{\rm vz}$ and $\gamma_{\rm lz}$ do {\it not} conform to the 
mean-field critical exponent (we instead find an exponent $\mu \approx 0.9$).}
\end{figure}

We now discuss the critical behavior of our equilibrium density profiles close 
to the vapor-zebra and liquid-zebra critical points. Since our DFT is a 
mean-field theory, we should recover mean-field critical exponents. This is not 
to suggest that the universality class of the AO model is the mean-field one -- 
it is not \cite{Vink2004} -- but rather that we wish to test the internal 
consistency of our theory. To this end, we introduce the vapor-zebra order 
parameter
\be
 \Delta_{\rm vz} = \frac{1}{\lambda} \int_0^\lambda
 \eta_{\rm c,z}(z) - \eta_{\rm c,v}(z) \, dz,
\ee
where $\eta_{\rm c,z}$ ($\eta_{\rm c,v}$) denotes the equilibrium colloid 
density profile of the zebra (vapor) phase. Note that $\Delta_{\rm vz}$ above is 
just the difference between the {\it average} colloid density of the vapor and 
zebra phase. The liquid-zebra order parameter is defined analogously
\be
 \Delta_{\rm lz} = \frac{1}{\lambda} \int_0^\lambda
 \eta_{\rm c,l}(z) - \eta_{\rm c,z}(z) \, dz.
\ee
Near the critical points, we expect power law decay of the order parameter
\be\label{eq:mf}
 \Delta_{\rm x} \propto t^\beta, \quad 
 t = \etapr - \eta_{\rm p,cr}^{\rm r,x} > 0, \quad
 \rm x \in (vz,lz),
\ee
with critical exponent $\beta=1/2$ for mean-field theory. We compute these order 
parameters as function of $\etapr$ and plot them on double logarithmic scales in 
\fig{fig:beta_RVPAR}, where on the horizontal axes the distance from the 
critical point~$t$ is shown. The power law of \eq{eq:mf} with mean-field 
exponent $\beta=1/2$ is strikingly confirmed!

Next, we consider the critical behavior of the vapor-zebra and liquid-zebra 
surface tension
\be\label{eq:mfst}
 \gamma_{\rm x} \propto t^\mu, \quad \rm x \in (vz,lz),
\ee
with $t>0$ defined as above, and where the critical exponent $\mu=3/2$ for 
mean-field theory. In \fig{fig:dft_st}, we plot both surface tensions as 
function of $t$, again using double logarithmic scales (this plot is simply a 
rescaling of the data of \fig{fig:tension_RVPAR}). For completeness, we also 
show the \LV surface tension of the {\it bulk} AO model, i.e.~in the absence of 
the laser field. The puzzling result is that, while the bulk tension conforms to 
$\mu=3/2$ as expected, we do {\it not} recover the expected mean-field critical 
exponent for the vapor-zebra and liquid-zebra surface tensions. From this we 
conclude that $\gamma_{\rm vz}$ and $\gamma_{\rm lz}$ do not become critical. We 
postulate there must be a \ahum{hidden} surface tension $\gamma_{\rm h}$ which 
instead conforms to \eq{eq:mfst} at the vapor-zebra and liquid-zebra critical 
points; finding the corresponding \ahum{hidden} surface is one of the challenges 
facing the simulations.

\section{Monte Carlo results}

We now use computer simulations to corroborate the DFT findings, and to shed 
light on the peculiar nature of the vapor-zebra and liquid-zebra critical 
points. We simulate the AO model (defined in \sect{sec:model}) inside the 
external potential of \eq{eq:laser} using grand canonical Monte Carlo 
\cite{Frenkel2001}. In the grand canonical ensemble, the colloid chemical 
potential $\muc$ and the polymer \ahum{reservoir packing fraction} $\etapr$ are 
fixed, while the number of colloids $\NC$ and polymers $\NP$ in the system 
fluctuate. We remind the reader that $\etapr$ plays the role of inverse 
temperature. To simulate efficiently, a grand canonical cluster move is used 
\cite{Vink2004}, combined with a biased sampling scheme \cite{Virnau2004}. The 
simulations are performed in a $V = L_x \times L_y \times L_z$ box with periodic 
boundaries. The laser field, \eq{eq:laser}, propagates along the edge $L_z$ of 
the box, and hence we choose $L_z = n \lambda$, with integer $n>0$, and 
$\lambda$ the wavelength of the field. In what follows, the colloid-to-polymer 
size ratio $q=1$, $\lambda=10$, and the laser field amplitude $V_0 / k_BT = 
0.4$. The key output of the simulations is the order parameter distribution 
$P(\etac)$ (OPD) defined as the probability to observe the system in a state 
with colloid packing fraction $\etac$. From the (normalized) OPD, one readily 
computes the average colloid packing fraction
\be
 \avg{\etac} = \int \etac \, P(\etac) d \etac,
\ee
as well as the colloidal compressibility
\be
 \chi_{\rm c} = V \left( \avg{\etac^2} - \avg{\etac}^2 \right),
\ee
and the Binder cumulant \cite{Binder1981}
\be
 Q \equiv \avg{m^2}^2 / \avg{m^4}, \quad m = \etac - \avg{\etac}.
\ee
We emphasize that the above quantities, as well as the OPD, depend on all the 
model parameters, in particular the system size, the imposed colloid chemical 
potential $\muc$, and the \ahum{inverse temperature} $\etapr$.

\subsection{Phase diagram}

\begin{figure}
\begin{center}
\includegraphics[width=0.9\columnwidth]{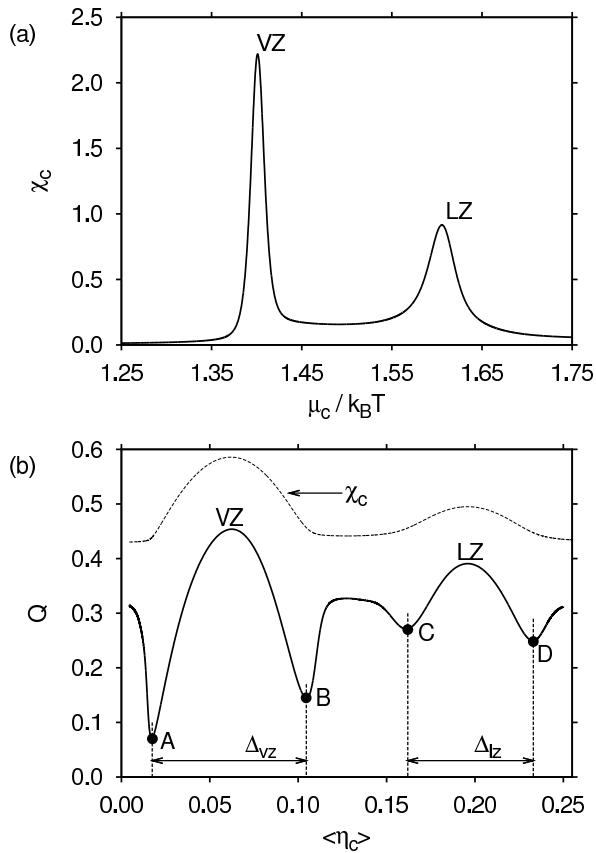}
\caption{\label{fig:chi} This figure illustrates how the vapor-zebra (VZ) and 
liquid-zebra (LZ) transitions are located in grand canonical simulations; 
$\etapr=1.0$, $L_x=L_y=12$, and $L_z=2\lambda=20$ are used in both plots. (a) 
The colloidal compressibility $\chi_{\rm c}$ versus the colloid chemical 
potential $\muc$; the chemical potential of the left (right) peak yields 
$\mucvz$ ($\muclz$). (b) The cumulant $Q$ as function of $\avg{\etac}$ (solid 
curve). The dashed curve shows the compressibility $\chi_{\rm c}$ on an 
arbitrary vertical scale. The compressibility maxima coincide with maxima in 
$Q$, the adjacent minima of which are labeled $A,B,C,D$. The horizontal 
double-arrows mark the vapor-zebra and liquid-zebra order parameters.}
\end{center}
\end{figure}

\begin{figure}
\begin{center}
\includegraphics[width=0.95\columnwidth]{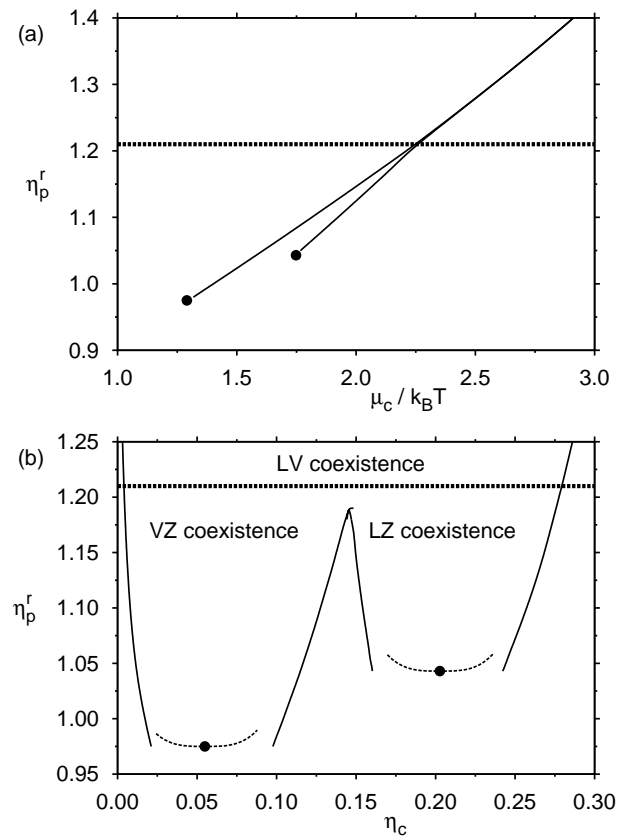}
\caption{\label{fig:pd_sim} Phase diagram of the AO model inside the laser field 
of \eq{eq:laser} in grand canonical representation (a) and reservoir 
representation (b); dots mark the critical points obtained via finite-size 
scaling. No scaling analysis was performed to locate the triple point, but based 
on the merging of the curves in (a), and also on the shape of the OPD 
(\fig{fig:tr}), we expect that $\etaprtr \sim 1.21-1.22$ (horizontal lines). The 
dashed curves in (b) are power laws corresponding to the critical exponent 
$\beta$ obtained in \fig{fig:fss}(b). As the system size is increased, the 
binodals obtained in the finite system (solid curves) smoothly approach these 
power laws. Also labeled in (b) are the various coexistence regions.}
\end{center}
\end{figure}

To obtain the phase diagram, we vary the colloid chemical potential $\muc$ at 
fixed \ahum{inverse temperature} $\etapr$; phase transitions correspond to peaks 
in the colloidal compressibility. An example is shown in \fig{fig:chi}(a), where 
$\chi_{\rm c}$ versus $\muc$ is plotted. We observe two sharp peaks, indicating 
two transitions. The left (right) peak corresponds to the vapor-zebra 
(liquid-zebra) transition, and from the peak position $\mucvz$ ($\muclz$) can be 
\ahum{read-off}. Note that, above the triple point ($\etapr>\etaprtr$), 
$\chi_{\rm c}$ versus $\muc$ reveals only one peak, then corresponding to a \LV 
transition. For a range of $\etapr$, we record the value(s) of the colloid 
chemical potential where $\chi_{\rm c}$ is maximal, and plot these as points in 
the $(\muc,\etapr)$~plane. The resulting phase diagram is shown in 
\fig{fig:pd_sim}(a), and the \ahum{inverted letter Y} topology predicted by the 
DFT is strikingly confirmed.

\begin{figure}
\begin{center}
\includegraphics[width=0.9\columnwidth]{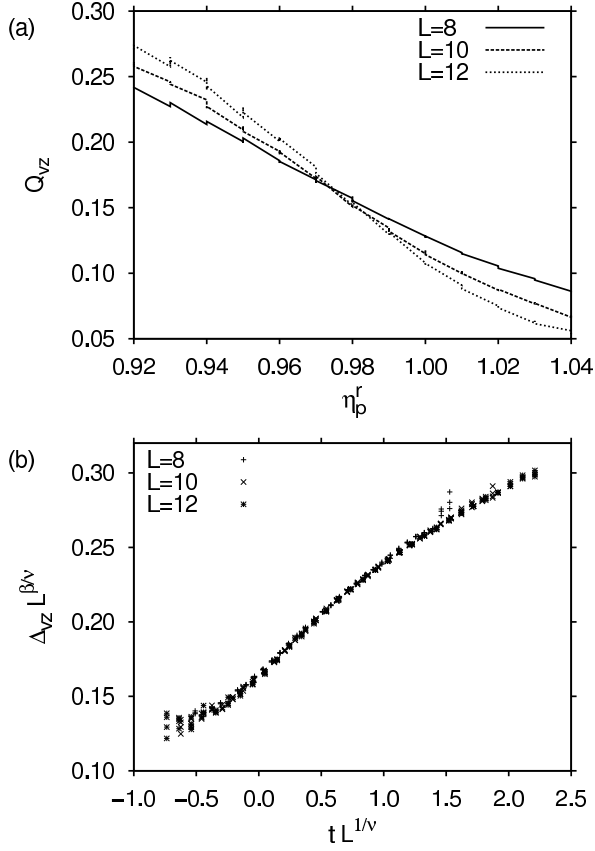}
\caption{\label{fig:fss} Finite-size scaling analysis of the vapor-zebra 
critical point. In both plots, $L_x = L_y = L$ is varied, while $L_z=2\lambda$ 
is fixed. (a) The cumulant $Q_{\rm vz}$ versus $\etapr$ for different $L$; the 
intersection yields $\etaprcrvz$. Note that, to the left (right) of the 
intersection, the cumulants approach $1/3$ ($0$), in agreement with \eq{eq:u1}. 
(b) Scaling plot of the vapor-zebra order parameter. Plotted is $\Delta_{\rm vz} 
L^{\beta/\nu}$ versus $t L^{1/\nu}$, where $\nu$ and $\beta$ were tuned until a 
good collapse of the data was observed.}
\end{center}
\end{figure}

The dots in \fig{fig:pd_sim} mark the vapor-zebra and liquid-zebra critical 
points, which we obtained using finite size scaling. For a given value of 
$\etapr$ and system size, we vary the colloid chemical potential $\muc$, and 
record the average colloid packing fraction $\avg{\etac}$, the colloidal 
compressibility $\chi_{\rm c}$, and the Binder cumulant~$Q$. A typical result is 
shown in \fig{fig:chi}(b), where $Q$ and $\chi_{\rm c}$ versus $\avg{\etac}$ are 
plotted (these curves are thus parametrized by $\muc$). The key message of 
\fig{fig:chi}(b) is that the compressibility maxima of the vapor-zebra and 
liquid-zebra transitions coincide with maxima in the cumulant. Adjacent to the 
cumulant maximum of the vapor-zebra transition, we observe two minima, indicated 
by the points
\be
 A \equiv \left( \avg{\etac}_{\rm vz}^-, \, Q^-_{\rm vz} \right), \quad
 B \equiv \left( \avg{\etac}_{\rm vz}^+, \, Q^+_{\rm vz} \right),
\ee
and, similarly, for the liquid-zebra transition
\be
 C \equiv \left( \avg{\etac}_{\rm lz}^-, \, Q^-_{\rm lz} \right), \quad
 D \equiv \left( \avg{\etac}_{\rm lz}^+, \, Q^+_{\rm lz} \right).
\ee
In the thermodynamic limit, it holds that \cite{Kim2005, Kim2003}
\begin{equation}\label{eq:u1}
 \lim_{L_x,L_y,L_z \to \infty} Q_{\rm vz} = \begin{cases}
 1/3 & \etapr<\etaprcrvz, \\
 0 & \etapr>\etaprcrvz,
 \end{cases}
\end{equation}
with $Q_{\rm vz} \equiv (Q^-_{\rm vz} + Q^+_{\rm vz})/2$, and $\etaprcrvz$ the 
value of $\etapr$ at the vapor-zebra critical point. Hence, by plotting $Q_{\rm 
vz}$ versus $\etapr$ for a number of different system sizes, curves for 
different system sizes intersect at $\etapr=\etaprcrvz$, which can be used to 
locate the critical point. Of course, to locate the liquid-zebra critical point, 
one analogously analyzes $Q_{\rm lz} \equiv (Q^-_{\rm lz} + Q^+_{\rm lz})/2$.

To perform the finite-size scaling analysis, we vary the lateral box extensions 
$L_x=L_y \equiv L$, but keep the elongated extension {\it fixed} at 
$L_z=2\lambda=20$. We assume that the divergence of the correlation length is 
\ahum{cut-off} in the $z$-direction by the laser field, and so we do not need to 
scale in this direction (this assumption will be justified in the next section 
where the static structure factor is discussed). Since the correlations diverge 
only in the two lateral directions, the critical behavior is effectively 
two-dimensional. In \fig{fig:fss}(a), we plot $Q_{\rm vz}$ versus $\etapr$ for 
three values of $L$. In agreement with \eq{eq:u1}, an intersection point is 
observed, from which we conclude that $\etaprcrvz \approx 0.975$. A similar 
analysis of the liquid-zebra transition yields $\etaprcrlz \approx 1.043$ (not 
shown). It is striking that the scaling analysis confirms the DFT prediction 
$\etaprcrlz>\etaprcrvz$. To estimate the colloid chemical potential $\mucvzcr$ 
of the vapor-zebra critical point in the thermodynamic limit, we measured 
$\mucvz$ of the compressibility maximum at $\etapr=\etaprcrvz$ for finite $L$, 
and extrapolated to $L \to \infty$ assuming $\mucvzcr - \mucvz \propto 1/L$. In 
this extrapolation, we ignore all subtleties concerning field and pressure 
mixing \cite{Kim2004}, but emphasize that such effects are tiny on the scale of 
the phase diagram in \fig{fig:pd_sim}(a). The resulting estimate reads as 
$\mucvzcr \approx 1.29$, while for the liquid-zebra transition $\muclzcr \approx 
1.75$ is obtained.

Next, we consider the scaling of the order parameter. The cumulant minima $A$ 
and $B$ of \fig{fig:chi}(b) readily yield $\Delta_{\rm vz} = \avg{\etac}_{\rm 
vz}^+ - \avg{\etac}_{\rm vz}^-$ as order parameter for the vapor-zebra 
transition. In the vicinity of the critical point $\Delta_{\rm vz} \propto 
t^\beta$, with $t = (\etapr - \etaprcrvz)/\etaprcrvz$, $t>0$, and critical 
exponent $\beta$. The result is shown in \fig{fig:fss}(b), where we used the 
standard finite-size scaling practice \cite{Newman1999} of plotting $\Delta_{\rm 
vz} L^{\beta/\nu}$ versus $t L^{1/\nu}$, with $\nu$ the correlation length 
critical exponent. Provided suitable values of $\etaprcrvz$, $\beta$ and $\nu$ 
are used, the data for different $L$ collapse. Reasonable collapses can indeed 
be realized, using for $\etaprcrvz$ the cumulant intersection estimate of 
\fig{fig:fss}(a), $\beta/\nu \sim 0.25-0.35$, and $\nu \sim 0.85-1.10$. An 
analysis of $\Delta_{\rm lz}$, which one obtains from the minima $C$ and $D$ of 
\fig{fig:chi}(b), yields similar results (not shown).

Finally, with the critical point parameters known, it becomes possible to 
calculate the phase diagram in {\it reservoir} representation, as is commonly 
done for the AO model. To this end, we record the colloid packing fraction of 
each of the cumulant minima $A,B,C,D$ in \fig{fig:chi}(b) as function of 
$\etapr$; the latter \ahum{trace-out} a curve (binodal) in the 
$(\etac,\etapr)$~plane. The result is shown in \fig{fig:pd_sim}(b), where dots 
again mark the critical points. To estimate the colloid packing fraction of the 
vapor-zebra critical point, we measured the finite-size \ahum{diameter} 
$\delta_{L, \rm vz} \equiv (\avg{\etac}_{\rm vz}^- + \avg{\etac}_{\rm vz}^+)/2$, 
using $\etapr=\etaprcrvz$ and the colloid chemical potential $\mucvz$ of the 
compressibility maximum; the diameter was then extrapolated to $L \to \infty$ 
assuming $\delta_{\infty,\rm vz} - \delta_{L,\rm vz} \propto 1/L$, which again 
ignores field and pressure mixing effects \cite{Kim2004}. In this way 
$\delta_{\infty,\rm vz} \approx 0.055$ is obtained, while an analogous procedure 
for the liquid-zebra critical point yields $\delta_{\infty,\rm lz} \approx 
0.203$.

\subsection{Nature of the critical point}

\begin{figure}
\begin{center}
\includegraphics[width=0.9\columnwidth]{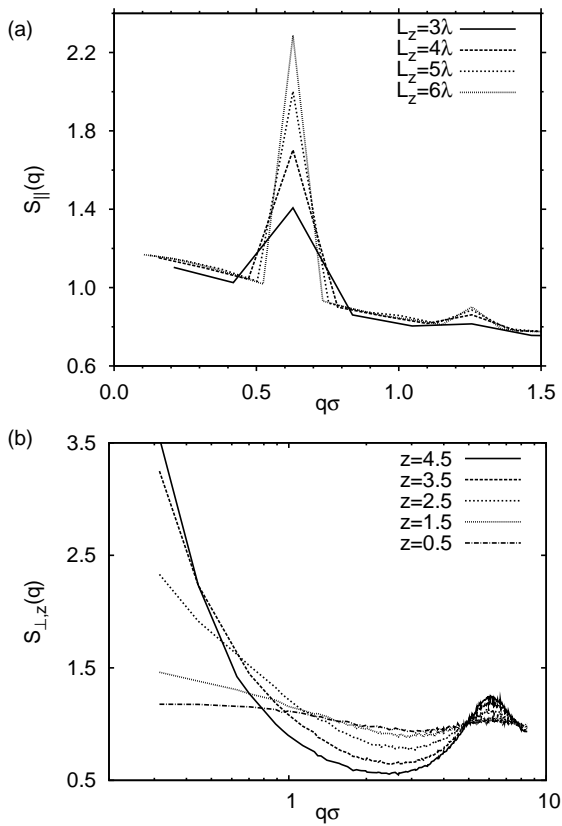}
\caption{\label{fig:sq} The colloid-colloid static structure factors obtained at 
the vapor-zebra critical point. (a) The structure factor $S_\parallel(q)$ 
measured in the direction of the laser field for several values of $L_z$, and 
fixed $L_x=L_y=10$. There is no divergence as $q \to 0$, indicating that 
critical fluctuations in the direction of the field are \ahum{cut-off}. (b) The 
structure factor $S_{\perp,z} (q)$ measured in directions perpendicular to the 
laser field for several values of $z$; the system size used equals $L_x=L_y=20$, 
and $L_z=4\lambda$. The key message is that $S_{\perp,z} (q)$ diverges as $q \to 
0$, but only for selected values of~$z$.}
\end{center}
\end{figure}

A key assumption in the finite-size scaling analysis of the previous section is 
that the critical correlations are \ahum{cut-off} in the $z$-direction, i.e.~the 
direction along which the laser field of \eq{eq:laser} propagates. To justify 
this assumption, we consider the colloid-colloid static structure factor 
$S(\vec{q}) = \avg{ \frac{1}{n_{\rm c}} | \sum_{j=1}^{n_{\rm c}} \exp(i \vec{q} 
\cdot \vec{r}_j)|^2}$, with $\avg{\cdot}$ a thermal average, the sum over all 
$j=1,\ldots,n_{\rm c}$ colloidal particles whose centers are inside a test 
volume~$v$, and $\vec{r}_j$ the position of the $j$-th colloid. As usual, 
wavevectors are given by $\vec{q} = 2\pi(k/L_x,l/L_y,m/L_z)$, integers $k,l,m 
\geq 0$, with the constraint that $k+l+m \neq 0$. We also introduce the 
wavevector magnitude $q^2=\vec{q} \cdot \vec{q}$.

To probe the correlations in the $z$-direction, we use as test volume~$v$ a 
narrow cylinder, with a diameter equal to the colloid diameter, placed parallel 
to the $z$-axis; due to the symmetry of the system, the location where the 
cylinder intersects the $xy$-plane is irrelevant. We then calculate the 
structure factor $S_\parallel(q)$, which is obtained using only the wavevectors 
$\vec{q}_\parallel \equiv 2\pi (0,0,m/L_z)$. In \fig{fig:sq}(a), we plot 
$S_\parallel(q)$ measured at the vapor-zebra critical point. These data were 
obtained in a semi-grand canonical ensemble: the colloid packing fraction and 
$\etapr$ are fixed to their critical values ($\etac = \delta_{\infty,\rm vz}, \, 
\etapr=\etaprcrvz$), while the number of polymers fluctuates. The important 
message to take from \fig{fig:sq}(a) is that, in the limit $q \to 0$, there is 
no sign of a divergence $S_\parallel(q) \to \infty$. Hence, there are no 
critical fluctuations in the $z$-direction. Note that the peak at $q \approx 
0.63$ corresponds precisely to $2 \pi / \lambda$ of the laser field. The 
analysis of $S_\parallel(q)$ at the liquid-zebra critical point leads to similar 
conclusions (not shown).

Next, we consider the static structure factor $S_{\perp,z} (q)$ measured in the 
lateral $xy$-directions, i.e.~perpendicular to the laser field. In this case, 
wavevectors take the form $\vec{q}_\perp \equiv 2\pi(k/L_x,l/L_y,0)$, and as 
test volume~$v$ we use a narrow $L_x \times L_y \times \Delta z$ slab, placed 
parallel to the $xy$-plane at \ahum{height}~$z$ (the slab thickness $\Delta z$ 
equals one colloid diameter). Since the system is not translation invariant in 
the $z$-direction, it matters at which $z$-coordinate the slab is located, and 
so $S_{\perp,z} (q)$ depends on $z$. In \fig{fig:sq}(b), we plot $S_{\perp,z} 
(q)$ at the vapor-zebra critical point for several $z$, again obtained using the 
semi-grand canonical ensemble. The key message to take from \fig{fig:sq}(b) is 
that $S_{\perp,z} (q)$ does diverge as $q \to 0$, but only for certain values 
of~$z$. The analysis of $S_{\perp,z} (q)$ at the liquid-zebra critical point 
leads to similar conclusions (not shown).

\begin{figure}
\begin{center}
\includegraphics[width=0.9\columnwidth]{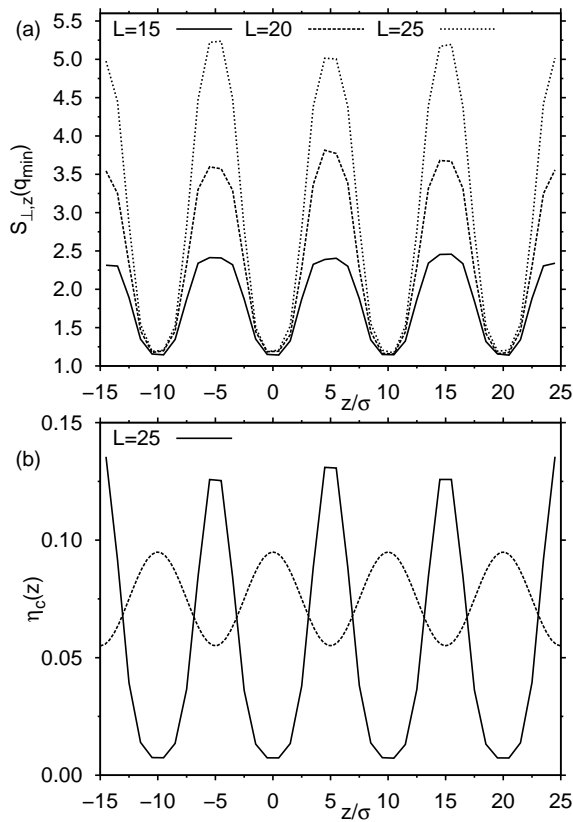}
\caption{\label{fig:prof_vz} Profiled quantities obtained at the vapor-zebra 
critical point. (a) The variation of $S_{\perp,z} (q_{\rm min})$ with $z$ for 
several $L$. (b) The average colloid packing fraction $\etac(z)$ measured along 
the $z$-direction (solid curve). The dashed curve shows the external laser 
potential of \eq{eq:laser} on an arbitrary vertical scale; regions dense in 
colloids coincide with minima of the potential.}
\end{center}
\end{figure}

\begin{figure}
\begin{center}
\includegraphics[width=0.9\columnwidth]{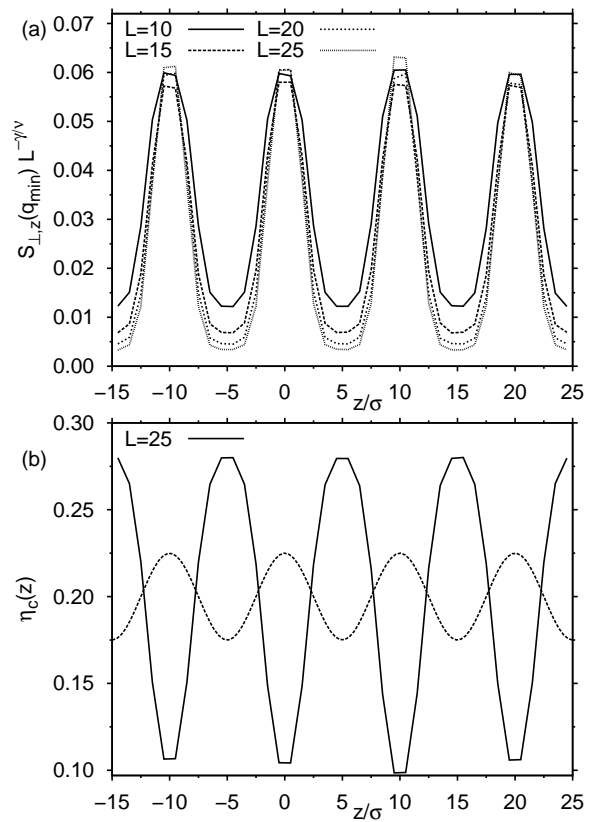}
\caption{\label{fig:prof_lz} The analogue of \fig{fig:prof_vz} for the 
liquid-zebra critical point. In (a), $\gamma/\nu=1.4$ is used.}
\end{center}
\end{figure}

{\it To summarize}: From the static structure factor $S_\parallel(q)$, we 
conclude that critical fluctuations in the $z$-direction are absent. This 
justifies our previous assumption that the critical behavior is effectively 
two-dimensional, such that finite-size scaling may be performed by varying only 
the lateral box extensions $L_x=L_y=L$, while keeping $L_z$ fixed. The analysis 
of $S_{\perp,z} (q)$ reveals that critical fluctuations indeed develop in the 
lateral directions, but only at certain $z$ values. The critical behavior is 
thus localized in effectively two-dimensional slabs perpendicular to the laser 
field, \ahum{sandwiched} between slabs where the system is non-critical. To make 
this explicit, we show in \fig{fig:prof_vz}(a) the variation of $S_{\perp,z} 
(q_{\rm min})$ with $z$ for the vapor-zebra critical point, where $q_{\rm min} = 
2\pi/L$ denotes the magnitude of the smallest accessible lateral wavevector. The 
figure strikingly shows that $S_{\perp,z} (q_{\rm min})$ diverges with $L$ only 
at selected $z$~values, i.e.~the critical behavior is indeed spatially localized 
in slabs. Note that the critical slabs correspond to regions of enhanced colloid 
density: $S_{\perp,z} (q_{\rm min})$ is \ahum{in-phase} with the colloid density 
profile $\etac(z)$ (\fig{fig:prof_vz}(b)). Interestingly, at the liquid-zebra 
critical point, this trend is reversed (\fig{fig:prof_lz}). Since $S_{\perp,z} 
(q_{\rm min}) \propto \chi_{\rm c}$ \cite{Rowlinson1982}, with $\chi_{\rm c}$ 
the colloidal compressibility, we expect $S_{\perp,z} (q_{\rm min}) \propto 
L^{\gamma/\nu}$ in the critical slabs \cite{Newman1999}. Here, $\gamma$ is the 
compressibility critical exponent; by fitting the peak values in 
\fig{fig:prof_vz}(a) to this scaling law, $\gamma/\nu \sim 1.3-1.4$ is obtained. 
For the liquid-zebra critical point, a similar ratio is found, see 
\fig{fig:prof_lz}(a), where $S_{\perp,z} (q_{\rm min}) L^{-\gamma/\nu}$ versus 
$z$ is shown (in this scaled representation, the peak values for different $L$ 
collapse). It is reassuring that the critical exponent ratios obtained in our 
scaling analysis conform to hyperscaling, $2\beta/\nu + \gamma/\nu=d=2$, as the 
reader can verify. Interestingly, our critical exponent ratios are rather 
different from 2D Ising values ($\beta/\nu=1/8, \gamma/\nu=7/4$), which we would 
naively have come to expect (only our $\nu$ estimate is somewhat consistent with 
$\nu=1$ of the 2D Ising model).

\subsection{The coexistence region}

\begin{figure*}
\begin{center}
\includegraphics[width=13cm]{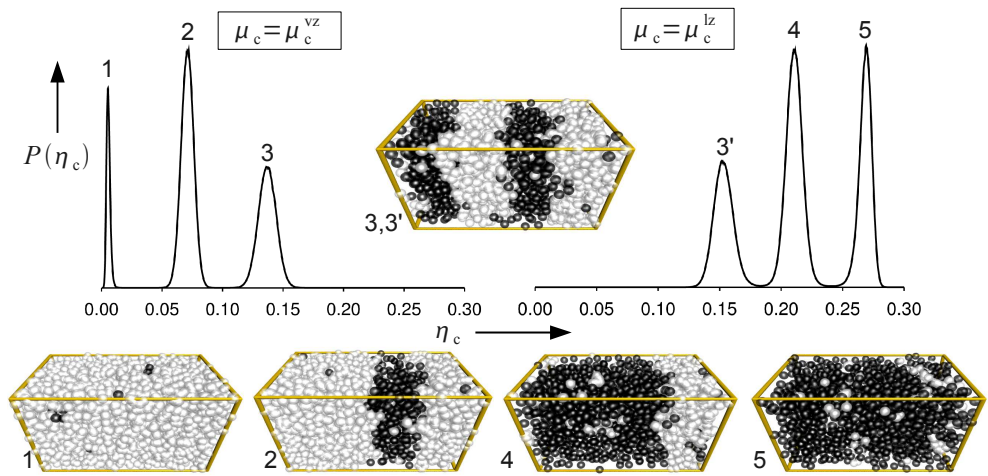}
\caption{\label{fig:opd} Analysis of the OPD $P(\etac)$ obtained at 
$\etapr=1.16$, which is above the critical points, but still below the triple 
point. The system size used is $L_z=2\lambda=20$, and $L_x=L_y=12$. Shown on the 
left is $P(\etac)$ at the vapor-zebra transition, while on the right $P(\etac)$ 
of the liquid-zebra transition is shown. The salient features are a number of 
peaks, whose meanings become clear upon inspection of snapshots. In the 
snapshots, the laser field is along the longest edge of the box; colloids are 
shown as black, polymers as white.}
\end{center}
\end{figure*}

We now consider the vapor-zebra and liquid-zebra two-phase coexistence regions, 
see \fig{fig:pd_sim}(b), where the corresponding transitions are first-order. To 
this end, we choose $\etapr$ above the critical points, but still below the 
triple point, and measure the OPD $P(\etac)$. In \fig{fig:opd}, we show 
$P(\etac)$ using $\muc=\mucvz$ of the vapor-zebra transition (left), and using 
$\muc=\muclz$ of the liquid-zebra transition (right). The striking feature is 
that the distributions reveal a number of peaks. We first discuss the OPD of the 
vapor-zebra transition. Here, the left peak~(1) reflects the vapor phase, 
i.e.~low colloid density, and high polymer density (see corresponding 
snapshot~1). Although not visible in the snapshot, we emphasize that the colloid 
density profile $\etac(z)$ of the pure vapor phase resembles that of 
\fig{fig:gas-fluid-zebra_GOETZE}, i.e.~there are (small) density modulations 
along the $z$-direction. The center peak~(2) corresponds to a mixed state, where 
a slab of vapor coexists with a slab of zebra phase (snapshot~2). Hence, a 
vapor-zebra interface is present, and the corresponding density profile 
$\etac(z)$ will schematically resemble \fig{fig:dft_cx}(b). Note that, due to 
periodic boundaries, the number of vapor-zebra interfaces is at least two. The 
right peak~(3) corresponds to a pure zebra phase (snapshot~3), with a density 
profile resembling the one shown in \fig{fig:gas-fluid-zebra_GOETZE}, 
i.e.~featuring large density oscillations. The meaning of the peaks in the OPD 
of the liquid-zebra transition follows analogously. In this case, peak~4 
reflects liquid-zebra coexistence, to be compared to the profile of 
\fig{fig:dft_cx}(c). Note that the density of the zebra phase at the vapor-zebra 
transition differs from that of the liquid-zebra transition.

\begin{figure*}
\begin{center}
\includegraphics[width=12cm]{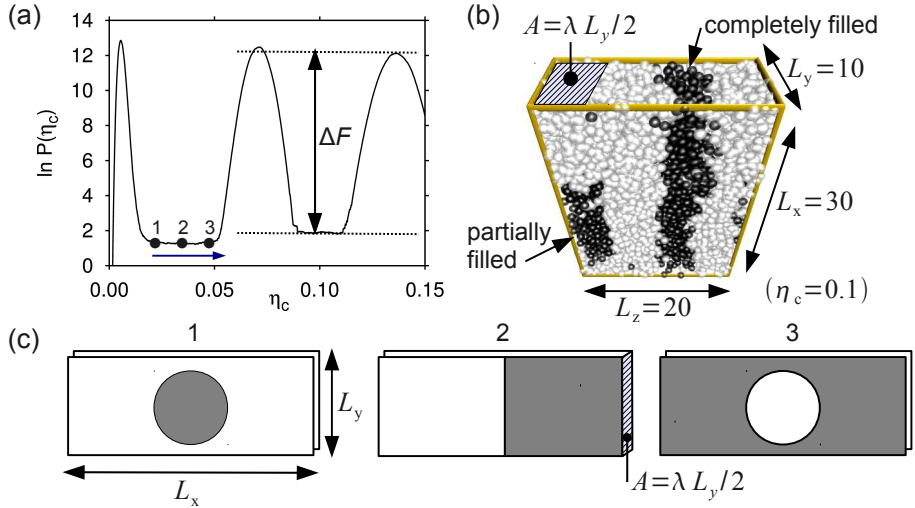}
\caption{\label{fig:cx} (a) The OPD obtained for $\etapr=1.16$ at the 
vapor-zebra transition; shown is $\ln P(\etac)$ versus $\etac$. The barrier 
$\Delta F$ reflects the interfacial free energy of the \ahum{hidden} interface 
(see details in text). (b) Snapshot taken at $\etac=0.1$, which is between the 
center and right peak of the OPD; the laser field propagates along $L_z$. 
Clearly visible is one completely filled slab, and one partially filled slab. 
The lateral area $A$ of the \ahum{hidden} interface is also indicated, where 
$\lambda$ is the wavelength of the field (colloids are shown as black, polymers 
as white). The schematic snapshots in (c) show the partially filled slab along 
the \ahum{path} $1 \to 2 \to 3$ of (a), where the $z$-direction is perpendicular 
to the plane of the paper (dark regions correspond to colloid-rich domains).}
\end{center}
\end{figure*}

Also of extreme interest are state points \ahum{between the peaks} in the OPD 
(\fig{fig:cx}). Here, we keep $\etapr=1.16$, but choose larger lateral box 
extensions, $L_x=30$ and $L_y=10$, while $L_z=2\lambda=20$. In \fig{fig:cx}(a), 
the logarithm of the OPD is shown, using $\muc=\mucvz$ of the vapor-zebra 
transition; note that $\ln P(\etac)$ may be regarded as {\it minus} the free 
energy of the system. As in \fig{fig:opd}, three peaks are visible: their 
meaning is the same as before. The snapshot of \fig{fig:cx}(b) was taken at 
$\etac=0.1$, which is between the center and right peak of the OPD. Again, 
vapor-zebra coexistence is observed, but the key difference with the coexistence 
state points of \fig{fig:opd} is that one of the periods of the field is only 
{\it partially} filled. Hence, in addition to a vapor-zebra interface 
perpendicular to the field, there is a smaller interface parallel to the field, 
indicated by the shaded area $A$. This is the \ahum{hidden} interface, whose 
presence was already implied by the DFT calculation. Note that $\ln P(\etac)$ 
around $\etac=0.1$ is essentially flat. Hence, once a partially filled slab has 
formed, it can be filled without any cost in free energy. This can be understood 
from the schematic snapshots of \fig{fig:cx}(c), which show top-down views 
(i.e.~looking along the $z$-direction) of the partially filled slab; the lateral 
area of the slab equals $L_x \times L_y$, while the slab thickness equals 
$\lambda/2$. The snapshots $1,2,3$ in (c) correspond to state points at the 
minimum between two peaks in the OPD, but with $\etac$ increasing from left to 
right (schematically resembling the \ahum{path} $1 \to 2 \to 3$ in 
\fig{fig:cx}(a)). In the first snapshot, a droplet of colloidal liquid has 
condensed. The droplet is cylindrical in shape; note that the area of the 
\ahum{hidden} interface in this configuration equals the circumference of the 
circle times the slab thickness. In the second snapshot, the droplet has grown 
so large it interacts with itself through the periodic boundaries, yielding two 
slab domains with two interfaces (the snapshot of \fig{fig:cx}(b) resembles this 
situation). Since $L_x \gg L_y$, the \ahum{hidden} interfaces form perpendicular 
to $L_x$; the shaded region $A$ marks the area of one of them. Since the free 
energy around the minimum of the OPD is flat, it follows that the interfaces do 
not interact \cite{Grossmann1993}, and so we obtain for the surface tension 
of the \ahum{hidden} interface \cite{Binder1981}
\be\label{eq:hid}
 \gamma_{\rm h} = \Delta F / 2A,
\ee
with $A = \lambda L_y/2$ (the factor $1/2$ in \eq{eq:hid} is a consequence of 
periodic boundaries, which lead to the formation of two interfaces). For 
$\etapr=1.16$, we obtain $\gamma_{\rm h} \approx 0.1 \, k_B T / \sigma^2$, which 
significantly exceeds the vapor-zebra surface tension. Finally, by increasing 
$\etac$ even further, one obtains the third snapshot, featuring a droplet of 
colloidal vapor. Note that \fig{fig:cx}(c) is just the \ahum{standard} droplet 
condensation transition in a two-dimensional system with periodic boundaries 
\cite{Fischer2010}.

\begin{figure}
\begin{center}
\includegraphics[width=1.0\columnwidth]{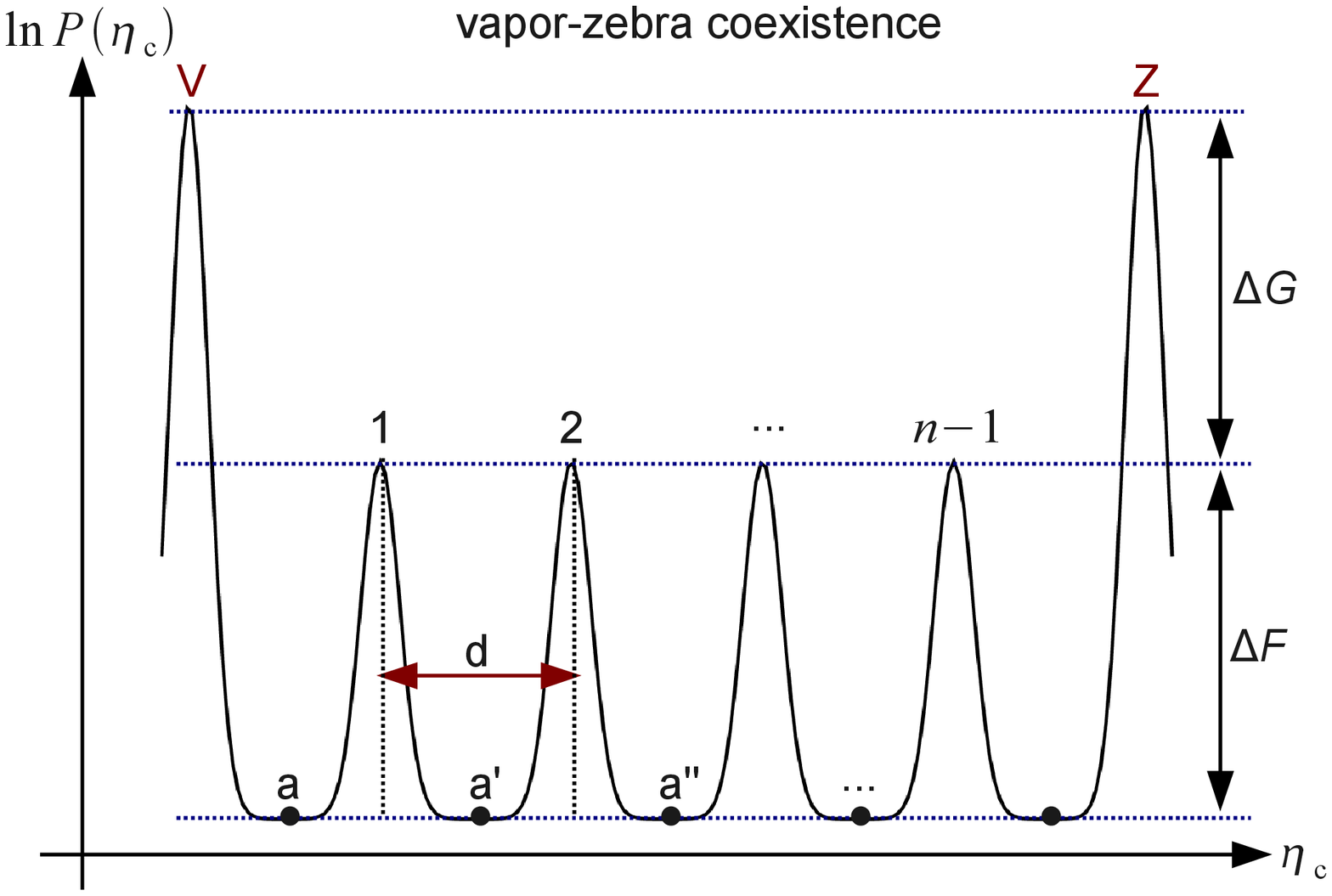}
\caption{\label{fig:id} Sketch of the logarithm of the OPD at the vapor-zebra 
transition, i.e.~using $\muc=\mucvz$, in a $L \times L \times L_z$ periodic box, 
$L_z=n\lambda$, and for $L$ large (solid curve). The two dominating peaks 
correspond to the pure vapor (V) and zebra phase (Z). The intermediate peaks 
$1,2,\ldots,n-1$ correspond to states where vapor and zebra coexist, with each 
period of the field {\it completely} filled with either one of the phases. The 
states $a,a',\ldots$ at the minima also correspond to vapor-zebra coexistence, 
but where one period of the field is only {\it partially} filled, as in 
\fig{fig:cx}(b). The barrier $\Delta G$ reflects the free energy cost of the 
vapor-zebra interface, $\Delta F$ that of the \ahum{hidden} interface; the 
respective scaling is given by \eq{eq:bar}.}
\end{center}
\end{figure}

Having understood the arrangement of the phases in the coexistence region, we 
expect the OPD at the vapor-zebra transition to scale with system size conform 
\fig{fig:id}. We assume a $L \times L \times L_z$ periodic box, $L_z = n 
\lambda$, with $L$ large. The two dominating peaks correspond to the pure vapor 
(V) and zebra (Z) phase. The intermediate peaks $1,2,\ldots,n-1$ correspond to 
states where vapor and zebra coexist, with each period of the field {\it 
completely} filled with either one of the phases. For each additional period of 
the field, one extra peak arises! The states $a,a',\ldots$ at the minima also 
correspond to vapor-zebra coexistence, but where one period of the field is {\it 
partially} filled, implying the presence of \ahum{hidden} interfaces 
(\fig{fig:cx}(b)). In the limit $L_z \to \infty$, one thus obtains an infinite 
sequence of intermediate peaks, separated by \ahum{distances} $d \propto 1/L_z$. 
The barrier $\Delta F$ reflects the free energy cost of the \ahum{hidden} 
interface; $\Delta G$ that of the vapor-zebra interface. As $L$ becomes large, 
we thus expect~\cite{Binder1981}
\be\label{eq:bar}
 \Delta F = \gamma_{\rm h} \lambda L, \quad
 \Delta G = 2 \gamma_{\rm vz} L^2.
\ee
In the thermodynamic limit $L \to \infty$, $\Delta G$ dominates: the 
intermediate peaks then become suppressed, and only the pure phase peaks (V,Z) 
remain. The OPD thus becomes bimodal, as it should since the transition is 
first-order between two phases \cite{Vollmayr1993}. The behavior of the OPD 
at the liquid-zebra transition follows analogously, although the precise values 
of $\gamma_{\rm h}$ will differ.

It now becomes clear why $\gamma_{\rm vz}$ and $\gamma_{\rm lz}$ cannot reveal 
critical behavior. The critical behavior was shown to be effectively 
two-dimensional, implying that the singular part of the interfacial free energy 
is due to {\it line tension}, i.e.~proportional to $L$. As \eq{eq:bar} shows, 
only the \ahum{hidden} interface reveals this required scaling. Consequently, 
$\gamma_{\rm h}$ becomes critical, while $\gamma_{\rm vz}, \gamma_{\rm lz}$ do 
not.

\begin{figure}
\begin{center}
\includegraphics[width=0.9\columnwidth]{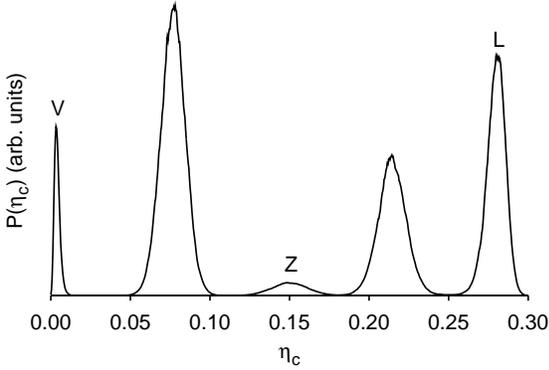}
\caption{\label{fig:tr} The OPD $P(\etac)$ at $\etapr=1.22$ which is close to 
the triple point; system sizes $L_x=L_y=8$, $L_z=2\lambda=20$ are used. The 
peaks corresponding to the pure vapor, zebra, and liquid phase are marked 
(V,Z,L), respectively.}
\end{center}
\end{figure}

\begin{figure}
\begin{center}
\includegraphics[width=0.9\columnwidth]{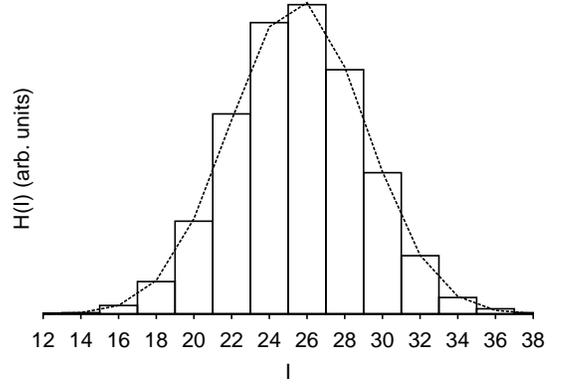}
\caption{\label{fig:int} Histogram $H(I)$ of the observed number of vapor-zebra 
interfaces $I$, obtained during a semi-grand canonical simulation inside the 
vapor-zebra coexistence region. The key message is that $I$ far exceeds two, 
showing that the tendency for macroscopic phase separation is weak. This is 
consistent with the DFT prediction that $\gamma_{\rm vz}$ is extremely small. 
The dashed curve shows the corresponding histogram for the (exactly known) 1D 
Ising model at infinite temperature.}
\end{center}
\end{figure}

Note that the OPDs for finite $L$ do not conform to \fig{fig:id}. For instance, 
in \fig{fig:opd}, the coexistence peaks (2,4) exceed those of the pure phases 
($1,3,3',5$). This is due to the extremely small values of $\gamma_{\rm vz}, 
\gamma_{\rm lz}$. For the system sizes $L$ accessible in our simulations, 
$\Delta G$ is essentially zero, meaning that the intermediate peaks are not 
suppressed. From the finite-size OPD, we thus obtain indirect confirmation of 
the DFT prediction that $\gamma_{\rm vz}, \gamma_{\rm lz}$ are extremely small. 
A second consequence is that the tendency of the system to macroscopically phase 
separate is weak. To test this assertion, we performed a semi-grand canonical 
simulation at $\etac=0.07$ and $\etapr=1.16$, using an extremely elongated box 
with $L_x=L_y=5$, $L_z=50\lambda$. The reader can verify in \fig{fig:pd_sim}(b) 
that this state point is deep inside the vapor-zebra coexistence region. 
Consequently, we expect macroscopic phase separation, implying the formation of 
$I=2$ vapor-zebra interfaces (since the system is periodic). In \fig{fig:int}, 
we have collected a histogram of observed $I$ values, obtained in a long 
simulation run. The key message is that the number of interfaces far exceeds 
two, providing further confirmation that $\gamma_{\rm vz}$ is small. In some 
sense, a $L \times L \times (L_z = n \lambda)$ system resembles a set of 
$i=1,\ldots,n$ slabs; to each slab we may assign a spin variable, say, $s_i=-1$ 
when the slab is filled with vapor, and $s_i=+1$ when filled with zebra (further 
justification of assigning spin variables $\pm 1$ to slabs follows from the DFT 
profiles of \fig{fig:dft_cx}(b), which show that the vapor-zebra interface is 
extremely sharp). When two neighboring slabs have different spin values, a 
vapor-zebra interface exists between them, which raises the free energy by an 
amount $\gamma_{\rm vz} L^2$. This is just the 1D Ising model \cite{footnote1}, 
with Hamiltonian ${\cal H}_{\rm Ising, 1D} = - J \sum_{i=1}^n s_{i-1} s_i$, $s_0 
\equiv s_n$, and coupling constant $J = \gamma_{\rm vz} L^2/2$. In the 
limit $L \to \infty$, one thus recovers the zero-temperature 1D~Ising model, and 
only here the system will macroscopically phase separate. However, due to the 
small value of $\gamma_{\rm vz}$ and the finite system size $L$, it is clear 
that our simulations are far removed from this limit, which also explains the 
result of \fig{fig:int}. In fact, the solid curve in \fig{fig:int} shows the 
distribution $H(I)$ for the 1D Ising model with $n=50$ spins and $J=0$ (with the 
constraint that the total magnetization $\sum_{i=1}^n s_i$ is zero).

Finally, we discuss the OPD at the triple point, where vapor, liquid, and zebra 
coexist. In the thermodynamic limit, the OPD becomes triple-peaked, each peak 
corresponding to one phase. In \fig{fig:tr}, we show the OPD near the triple 
point for a finite system. We indeed observe all three phases~(V,Z,L) 
simultaneously, but the coexistence peaks are still profoundly present. This 
once more confirms the extremely low values of $\gamma_{\rm vz}, \gamma_{\rm 
lz}$, even near the triple point, where they are maximal. To describe the 
coexistence in terms of spin variables, as done above, now requires 3-state 
spins, which might induce 1D 3-state Potts behavior at the triple point (this 
could be an interesting topic for further study). Above the triple point, only 
vapor and liquid can coexist, and here the OPD is bimodal again \cite{Vink2004}. 
In the \LV coexistence region, we could again map the system onto the 1D Ising 
model, but with coupling constant $J = \gamma_{\rm lv} L^2/2$. We have 
verified that, due to the substantially larger value of $\gamma_{\rm lv}$, the 
tendency of the system to phase separate is now much stronger. Of course, for 
the {\it bulk} AO model, the mapping onto the 1D Ising model does not apply (in 
this case, the external field, \eq{eq:laser}, which ultimately supplies the 
underlying 1D lattice structure, is absent).

\section{Conclusions}

In conclusion, we have studied fluid phase separation inside a static 
one-dimensional oscillatory external field. The actual DFT calculations and 
simulations were performed for the Asakura-Oosawa model of colloid-polymer 
mixtures, but we expect that our findings will apply to any three-dimensional 
fluid with a bulk \LV critical point. As was already established in a previous 
work \cite{Gotze2003}, the external field \ahum{splits} the bulk critical point 
into two new critical points, and one triple point. This leads to a phase 
diagram with three coexistence regions, featuring (1) vapor-zebra coexistence, 
(2) liquid-zebra coexistence, and (3) \LV coexistence. All three phases (vapor, 
liquid, zebra) are characterized by density modulations along the field 
direction, but the modulations are most pronounced in the zebra phase. The 
improved DFT calculation of the present work shows that the temperatures of the 
two critical points differ slightly from each other. In addition, we calculated 
the surface tensions associated with all three coexistence regions, and found 
the vapor-zebra and liquid-zebra tensions to be extremely small. The DFT 
calculation also reveals that the latter surface tensions do not yield the 
expected mean-field critical exponents (even though our DFT is a mean-field 
theory).

Computer simulations and finite-size scaling confirm all the trends predicted by 
the DFT. The reason that the vapor-zebra and liquid-zebra tensions do not show 
critical behavior is due to the fact that the external field divides the system 
into a series of effectively two-dimensional slabs, stacked on top of each other 
along the field direction. The critical correlations diverge only in directions 
perpendicular to the field, and the corresponding surface tension is one arising 
from phase coexistence inside single slabs. A surprising finding is that the 
critical behavior is confined to certain slabs only; depending on the critical 
point, either the low or high density slabs become critical. Along the field 
direction, and above the critical points, the arrangement of slabs can be 
conceived as a one-dimensional Ising chain, at effectively zero temperature in 
the thermodynamic limit, whereby each slab represents one Ising spin variable. 
Hence, there will be macroscopic phase separation in all three coexistence 
regions, but only in the limit where the lateral extensions $L$ of the system 
become large. Regarding the vapor-zebra and liquid-zebra coexistence regions, 
the tendency to phase separate is particularly weak, due to the extremely low 
vapor-zebra and liquid-zebra surface tensions. According to our DFT 
calculations, the latter tensions are of the order $10^{-6} \, k_BT / \sigma^2$. 
This value is too low to be quantitatively measured in simulations. However, the 
weak tendency of the system to macroscopically phase separate, as observed in 
our simulations, does confirm that the latter surface tensions must be extremely 
small.

Our results could be verified in real-space experiments of colloid-polymer 
mixtures using, for instance, confocal microscopy \cite{Vossen2004}. In fact, 
{\it bulk} criticality in these systems has already been analyzed in this manner 
\cite{Royall2007}. The inclusion of a standing optical field appears to be a 
feasible extension \cite{Freire1994, Jenkins2008}. In the presence of such a 
field, the much weaker tendency of the system to macroscopically phase separate 
should be easily detectable.

A remaining puzzle is why our finite-size scaling analysis does not reveal 
two-dimensional Ising critical exponents. Of course, the division of the system 
into slabs is not absolute: particles can still diffuse between slabs. Perhaps 
this modifies the universality class, but the underlying theoretical mechanism 
remains yet to be elucidated \cite{diehl}. We are currently planning simulations 
of the lattice Ising model to address these issues (the simplicity of the latter 
model probably allows for a more accurate scaling analysis using larger system 
sizes). It would also be interesting to extent the analysis to external 
potentials more complicated than the one of \eq{eq:laser}. Examples include a 
superposition of several waves resulting in two-dimensional periodic 
\cite{Fischer2011, Bechinger2001, Reichhardt2002, ElShawish2008, Lowen2009, 
Franzrahe2009} or quasi-crystalline patterns \cite{Mikhael2010, 
Schmiedeberg2008}. The phase diagram of a system close to its bulk critical 
point inside these confining potentials still needs to be explored. Again, the 
question is whether new critical points arise, and to what extent the emerging 
critical behavior is affected by the details of the confining potential.

\begin{acknowledgments}

We thank H. W. Diehl for helpful discussions. This work is financially supported 
by the SPP~1296 program, the SFB~TR6, and the Emmy Noether program (VI~483/1-1) 
of the {\it Deutsche Forschungsgemeinschaft}.

\end{acknowledgments}


\appendix

\section{Density functional theory background}
\label{app:dft}

The main variables in our density functional theory are the one-body densities 
$\rho_i(\mathbf{r})$ of colloids ($i=\rm c$) and polymers ($i=\rm p$), which 
describe the microscopic behavior of the system for a given set of parameters 
(temperature $T$ and chemical potential $\mu_i$). Based on the existence proof 
\cite{Mermin1965} that there is a grand canonical free energy functional 
$\Omega(T,\muc,\mup,[\rhoc,\rhop])$ which gets minimal for the equilibrium 
density, we use the fundamental measure approach \cite{Schmidt2000} to 
approximate this functional. The grand canonical free energy functional of a 
colloid-polymer mixture in a three-dimensional system can be split as
\begin{equation*}
\begin{split}
&\Omega[\rhoc(\mathbf{r}),\rhop(\mathbf{r})]= \sum_{i= {\rm c,p}}\mathcal{F}_{\rm id}[\rho_i(\mathbf{r})] \quad + \\
&\sum_{i= {\rm c,p}}\int \text{d}\mathbf{r} \rho_i (\mathbf{r}) \left[ V_{\text{ext},i}(\mathbf{r})-\mu_i\right] 
+\mathcal{F}_{\rm exc}[\rhoc(\mathbf{r}),\rhop(\mathbf{r})],
\end{split}
\end{equation*}
with the external potential $V_{\text{ext},i}(\mathbf{r})$ acting on component 
$i$, keeping in mind a general description where both colloids and polymers are 
inside an external field. We use \eq{eq:laser} for the external potential 
$V_{\rm ext,c}$ acting on the colloids, and set the external potential acting on 
the polymers to zero: $V_{\rm ext,p}=0$. In the above, 
$\mathcal{F}_\text{id}[\rho_i(\mathbf{r})]$ is the free energy of an ideal gas
\begin{equation*}
 \mathcal{F}_\text{id}[\rho_i(\mathbf{r})] =
 k_BT\int \text{d}\mathbf{r} \rho_i (\mathbf{r})  \left[\ln\left(\rho_i (\mathbf{r})\Lambda_i^3\right)-1\right],
\end{equation*}
including the (irrelevant) thermal wavelength $\Lambda_i$ of the particles of 
species $i$, an external energy part and the nontrivial excess free energy 
$\mathcal{F}_{\rm exc}[\rhoc(\mathbf{r}),\rhop(\mathbf{r})]$, which results from 
the interactions of the particles. We approximate this excess free energy 
functional as the integral of a free energy density $\Phi(\{n_\nu^{\rm 
c}(\mathbf{x})\},\{n_\gamma^{\rm p}(\mathbf{x})\})$ as
\begin{equation*}
 \mathcal{F}_{\rm exc}[\rhoc(\mathbf{r}),\rhop(\mathbf{r})] =
 k_BT \int \text{d}\textbf{r}~\Phi(\{n_\nu^{\rm c}(\mathbf{x})\}, \{n_\gamma^{\rm p}(\mathbf{x})\}),
\end{equation*}
depending on weighted densities $n_\nu^i(\mathbf{x})$ given by the convolution 
of the actual density profiles with weight functions
\begin{equation*}
 n_\nu^i(\mathbf{x})=\int \text{d}\mathbf{r}\rho_i(\mathbf{r})w_\nu^i(\mathbf{x}-\mathbf{r}).
\end{equation*}
The set of weight functions (which are independent of the density profiles) is 
given by
\begin{flalign*}
 &w_3^i(\textbf{r})=\theta\left(\frac{\sigma_i}{2}-r\right), 
 w_2^i(\textbf{r})=\delta \left(\frac{\sigma_i}{2}-r\right), \nonumber \\
 &w_1^i(\textbf{r})=\frac{1}{2\pi \sigma_i}\delta \left(\frac{\sigma_i}{2}-r\right), 
 w_0^i(\textbf{r})=\frac{1}{2\pi \sigma_i^2}\delta \left(\frac{\sigma_i}{2}-r\right), \nonumber \\
 &\textbf{w}_2^i(\textbf{r})=\delta \left(\frac{\sigma_i}{2}-r\right)\frac{\textbf{r}}{r}, 
 \textbf{w}_1^i(\textbf{r})=\frac{1}{2\pi \sigma_i}\delta \left(\frac{\sigma_i}{2}-r\right)\frac{\textbf{r}}{r}, \nonumber \\
 &\hat{\textbf{w}}_2^i(\textbf{r})=\delta \left(\frac{\sigma_i}{2}-r\right)\left[\frac{\textbf{rr}}{r^2}-\frac{\hat{\textbf{1}}}{3}\right],
\end{flalign*}
with $r=|\mathbf{r}|$, the step function $\theta(r)$, the Dirac function 
$\delta(r)$, and the identity matrix $\hat{\textbf{1}}$. The weight functions 
are of different tensorial rank, i.e.~scalars $w_3^i,w_2^i,w_1^i,w_0^i$, vectors 
$\textbf{w}_2^i,\textbf{w}_1^i$, and a second rank tensor 
$\hat{\textbf{w}}_2^i$. The excess free-energy density 
$\Phi=\Phi_1+\Phi_2+\Phi_3$ is written as
\begin{align*}
 \Phi_1&=\sum \limits_{i={\rm c,p}}n_0^i\varphi^i(n_3^{\rm c},n_3^{\rm p}), \nonumber \\
 \Phi_2&=\sum \limits_{i,j={\rm c,p}}\left(n_1^i n_2^j-\mathbf{n}_1^i\cdot \mathbf{n}_2^j\right)\varphi^{ij}(n_3^{\rm c},n_3^{\rm p}), \nonumber \\
 \Phi_3&=\frac{1}{8\pi}\sum \limits_{i,j,k={\rm c,p}}\left(\frac{n_2^i n_2^j n_2^k}{3}-n_2^i\mathbf{n}_2^j\cdot \mathbf{n}_2^k\right. \nonumber \\ 
 &~~+\left.\frac{3}{2}\left[\mathbf{n}_2^i\hat{\mathbf{n}}_2^j\mathbf{n}_2^k-\text{Tr}\left(\hat{\mathbf{n}}_2^i\hat{\mathbf{n}}_2^j\hat{\mathbf{n}}_2^k\right)\right]\right)\varphi^{ijk}(n_3^{\rm c},n_3^{\rm p}),
\end{align*}
where $\varphi^{i\ldots k}(\eta^{\rm c},\eta^{\rm p})=\beta 
\frac{\partial^m}{\partial \eta^i\ldots \partial \eta^k}F_{\text{0D}}(\eta^{\rm 
c},\eta^{\rm p})$ denotes the derivatives of the 0D free energy $\beta 
F_\text{0D}(\eta^{\rm c},\eta^{\rm p})=(1-\eta^c-\eta^p)\ln(1-\eta^{\rm 
c})+\eta^{\rm c}$. We obtain the equilibrium density profiles 
$\rho_i^{(0)}(\mathbf{r})$ by minimizing the Gibbs free energy functional,
\begin{equation*}
 \left. \frac{\delta \Omega[\rhoc(\mathbf{r}),\rhop(\mathbf{r})]}{\delta 
\rho_i(\mathbf{r})}\right|_{\rho_i(\mathbf{r})=\rho_i^{(0)}(\mathbf{r})}=0.
\end{equation*}
This yields the Euler-Lagrange or stationarity equations 
\begin{equation*}
 \int \text{d}\mathbf{x}\frac{\delta \Phi}{\delta \rho_i(\mathbf{r})} + 
 \ln(\Lambda_i^3\rho_i(\mathbf{r})) + \beta V_{\text{ext},i}(\mathbf{r})-\beta \mu_i=0.
\end{equation*}
By inserting the equilibrium profiles into the functional, we obtain the grand 
canonical free energy and can thus calculate phase diagrams and interfacial
properties.

\newpage

\end{document}